\def   \ni {\noindent}

\def   \ssk {\vskip  5truept}

\def   \bsk {\vskip 15truept}

\def   \newline {\hfil\break}

\documentstyle[epsfig]{article}

\def\ergscm2{~{\rm erg}~{\rm s}^{-1}~{\rm cm}^{-2} }
\def\TeV{~\rm{TeV}}
\def\GeV{~\rm{GeV}}
\def\MeV{~\rm{MeV}}

\def\K{~\rm{K}}
\def\G{~\rm{G}}

\def\ms{~\rm{ms}}

\newbox\grsign \setbox\grsign=\hbox{$>$} \newdimen\grdimen \grdimen=\ht\grsign
\newbox\simlessbox \newbox\simgreatbox \newbox\simpropbox
\setbox\simgreatbox=\hbox{\raise.5ex\hbox{$>$}\llap
     {\lower.5ex\hbox{$\sim$}}}\ht1=\grdimen\dp1=0pt
\setbox\simlessbox=\hbox{\raise.5ex\hbox{$<$}\llap
     {\lower.5ex\hbox{$\sim$}}}\ht2=\grdimen\dp2=0pt
\setbox\simpropbox=\hbox{\raise.5ex\hbox{$\propto$}\llap
     {\lower.5ex\hbox{$\sim$}}}\ht2=\grdimen\dp2=0pt
\def\simgreat{\mathrel{\copy\simgreatbox}}

\begin{document}

\hsize 5truein
\vsize 8truein
\font\abstract=cmr8
\font\keywords=cmr8
\font\caption=cmr8
\font\references=cmr8
\font\text=cmr10
\font\affiliation=cmssi10
\font\author=cmss10
\font\mc=cmss8
\font\title=cmssbx10 scaled\magstep2
\font\alcit=cmti7 scaled\magstephalf
\font\alcin=cmr6 
\font\ita=cmti8
\font\mma=cmr8
\def\ref{\par\noindent\hangindent 15pt}
\null
%\vskip 3.0truecm
%\baselineskip = 12pt

% beginning of font "title"

\title{\ni Broad-Band Model Spectra of Gamma-Ray Emission\\
from Millisecond Pulsars}
\bsk\bsk
\author{\ni T. Bulik$^1$ and B. Rudak$^2$}
\bsk
\affiliation{\ni Nicolaus Copernicus Astronomical Center\\
     $^1$ Bartycka 18, 00-716 Warsaw, Poland\\
     $^2$ Rabia{\'n}ska 8, 87-100 Toru{\'n}, Poland}

\bsk
\baselineskip = 12pt

\abstract{ABSTRACT \ni

We present  spectra of pulsed gamma-ray emission expected
from millisecond pulsars within the framework of polar-cap
models.  The spectra are a superposition of three components due
to curvature (CR), synchrotron (SR), and Compton upscattering
(ICS) processes. The CR component dominates  below $100\,$GeV
and the ICS component exhibits a peak at $\sim 1\,$TeV.
The CR component should be observable from J0437-4715 with
the next generation gamma-ray telescopes, like e.g. GLAST. }

\bsk
\baselineskip = 12pt


\bsk
\baselineskip = 12pt

% beginning of font "text"

\text{\ni 1. INTRODUCTION
\ssk
\ni     

According to model predictions (Sturner \& Dermer 1994, Rudak \&
Dyks 1998a)  the sensitivity of present-day satellite gamma-ray
experiments like EGRET is about one order of magnitude too low
for millisecond pulsars with high spin-down fluxes to be
detected; for upper limits from EGRET see Nel et~al.~(1996). 
However, objects like J0437-4715 should be accessible
with future missions like {GLAST}, thus  providing a testing
ground for  magnetospheric models relevant for millisecond
pulsars.  The aim of this paper is to present major spectral
features of pulsed gamma-ray emission (for $E \simgreat 1\MeV$) 
expected when cascades of secondary particles,  triggered by
ultrarelativistic primary electrons,  develop above polar caps
of millisecond pulsars with dipolar magnetic fields around
$10^9\G$.  

\begin{figure}[h]
\centerline{\psfig{angle=-90,file=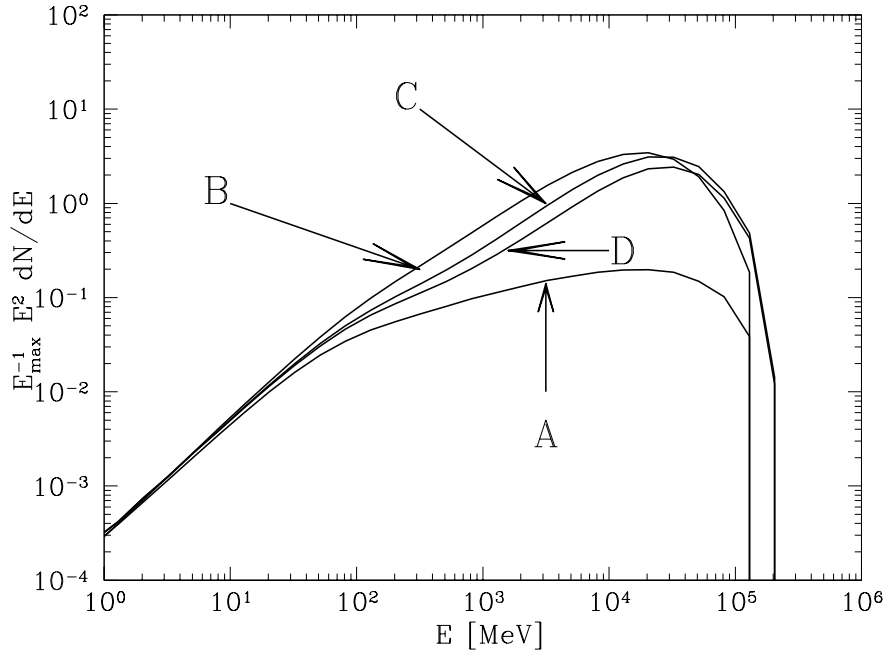, width=12cm}}
\caption{FIGURE 1. The CR energy spectral component due to a single primary
electron
injected  at the outer rim of a polar cap of the standard
millisecond pulsar. We normalise the
spectrum with
$E_{\rm max}$ - maximum  energy of the electron, mediated
between acceleration and cooling rates (models B, C, and D). For
model A (instant acceleration at the cap surface), $E_{\rm max}$
is equal to the initial energy $E_0 = 10^7\MeV$. Pulsar
parameters are
$B_{\rm pc} = 10^9\G$, and $P = 3 \ms$. }
\end{figure}

\bsk
\ni 3. ACCELERATION OF PRIMARY ELECTRONS 
\ssk
\ni 

We investigated four different models of electron acceleration.
Electrons are instantly accelerated to ultrarelativistic energy
in model~A, i.e. they are injected with $E_{\rm init} = E_{\rm
max}$.  The values of $E_{\rm init}$ were chosen (see Rudak \&
Dyks 1998b for details) to be just above the threshold value for
creation of Sturrock pairs: for a standard millisecond pulsar
($B_{\rm pc} =10^9\G, \, P = 3\ms$) $E_{\rm init} = 1.07\times
10^7\MeV$. For J0437-4715, $B_{\rm pc} =7.4\times 10^8\G, \, P =
5.75\ms$, and  $E_{\rm init} = 1.54\times 10^7\MeV$, where we
used the magnetic field  derived from kinematically corrected
$\dot P$ by Camillo et~al. (1994). For comparison, three other
models (B, C, and D) for particle  acceleration were considered,
with different dependencies  of the longitudinal electric
field  on  the height $h$:
\begin{equation}
{\cal E} (h) = {V_0\over r_{\rm pc}} \times \cases{ \exp\lbrack-{h\over r_{\rm pc}}\rbrack, &model B\cr
\exp\lbrack -\left({h\over r_{\rm pc}}\right)^2\rbrack, &model C \cr
\mathrm{max} \left\{\left[ 1 -\left({h\over r_{\rm
pc}}\right)^2\right],
0\right\}, &model D \cr}
\label{efield}
\end{equation}
The characteristic scale height of the electric field is equal
to the polar cap radius $r_{\rm pc}$ in each case. For each
pulsar, the value of the potential drop $V_0$ was chosen to
yield a similar number of pairs as in the model A.

\begin{figure}[h]
\centerline{\psfig{angle=-90,file=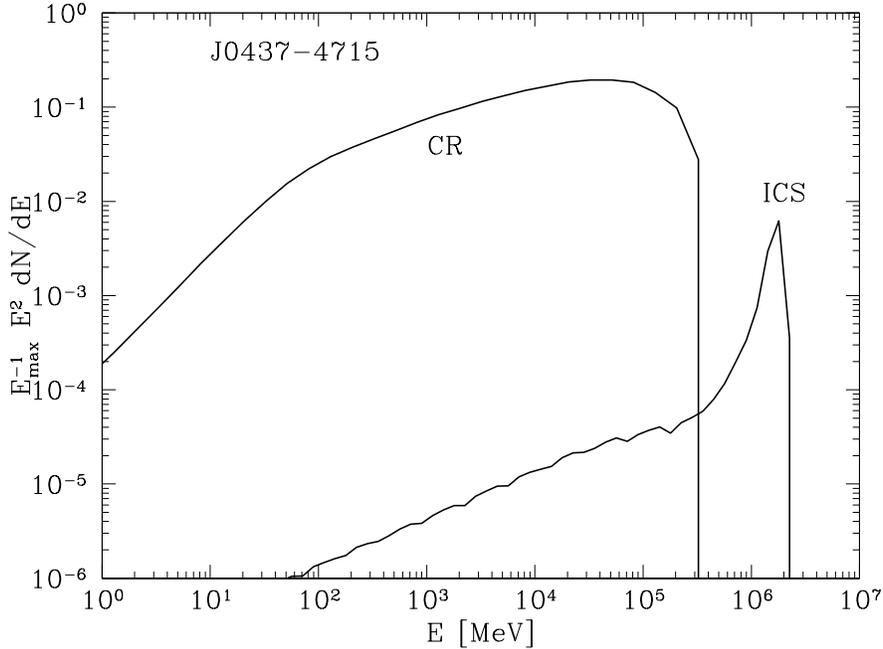, width=12cm}}
\caption{FIGURE 2. Model energy spectrum of J0437-4715 per primary
electron. We use
model A with $E_{\rm init}=1.54\times 10^7\,$MeV, and soft
photons
coming from the surface with the temperature $4\times 10^5\K$
(Edelstein et~al. 1995). The labels CR and ICS denote 
the curvature and inverse Compton components respectively.
Each component includes transfer effects due to magnetic absorption
and subsequent synchrotron radiation of the pairs.}
\end{figure}

\bsk
\ni 3. PRINCIPAL SPECTRAL COMPONENTS 
\ssk
\ni

{\bf Curvature Radiation.} CR is the dominant spectral
component  in the energy range from $1\MeV$ out to a cutoff at
$\simgreat 0.1\TeV$. Its characteristic features are shown in
Figure 1 for the different acceleration models.  Due to
smaller   curvature radii implying higher values of the
characteristic photon energy, and also to much weaker magnetic
absorption ($\gamma B \rightarrow e^\pm$), the high-energy
cutoff is far beyond the cutoffs expected for classical,
high-$B$ pulsars (Dyks \& Rudak 1998).  
The slope $\alpha$ of radiation energy per logarithmic energy
bandwith ($E^2 dN/dE \propto E ^{-\alpha}$) below $100\MeV$ is
insensitive to the model of acceleration, and is close to $4/3$.
Above the break at $\sim 100 \MeV$ the spectrum levels off to a
degree which depends on the acceleration model.  For models A
and B the slope between $100\MeV$ and $10\GeV$ is $1/3$ and
$0.84$, respectively. For models C and D the spectra are
practically indistinguishable, with slopes  approximately equal to $
0.7$ between $100\MeV$ and $1\GeV$ and then steepening to
about $0.87$ between $1\GeV$ and $10\GeV$.

{\bf Non-magnetic Inverse Compton.} For millisecond pulsars,
magnetic fields at the surface are much weaker than the critical
field $B_{\rm crit} = 4.414\times 10^{13}\G$. Thus the
scattering  process of electrons in a field of ambient photons
is well described by the Klein-Nishina relativistic non-magnetic
cross section. Our treatment of the scattering process follows
Daugherty \& Harding (1989).  In the non-magnetic case the
optical depth for scattering of an electron is small. Thus
scattering is not a relevant electron decelaration process. Most
of the scatterings take place when the electron is at the height
of the order of the size of soft-photons source. We consider two
cases of the geometry of the soft photon source. In the first
one the soft  photons originate  on a hot  polar cap  with
temperature  $\simgreat 10^6\,$K.  In this case most of the ICS
photons will pair produce and will not be able to escape.  In
the second scenario  the soft photons come from the entire
surface of the neutron star, with the temperature  around  a few
times $10^5\,$K. Here most of the scattering events take place
at the distance of a few neutron star radii from the surface,
thus most  photons will escape freely.

\bsk
\ni 4. DISCUSSION 
\ssk
\ni

We have calculated broad-band  spectra of non-thermal origin for
millisecond pulsars within the framework of  polar-cap models
with magnetospheric activity induced by curvature radiation of
beam particles.   This non-thermal emission is a superposition
of curvature, inverse Compton and   synchrotron radiation.  The
curvature component dominates the region between 1\,MeV and
100\,GeV. The slope of the spectrum  in the range between 
$100\,$MeV and $10\,$GeV is sensitive to the details of the
electron acceleration process. The synchrotron component becomes important 
only below $\sim 1\MeV$.
The ICS component has a narrow
peak around $1\,$TeV. In the case of the nearby  millisecond
pulsar J0437-4715 the curvature component should be detectable
with the next generation gamma-ray telescopes (e.g. GLAST). 
The ICS component  is about two orders of magntitude
too weak for present day ground based Cherenkov detectors (Bulik
\& Rudak 1998).  The expected broad band gamma-ray spectrum of
this pulsar is shown in Figure~2. 

}

\bsk
\baselineskip = 12pt
{\abstract \ni ACKNOWLEDGMENTS
This research was funded by the KBN grant 2P03D00911.
BR is grateful for travel support from KBN grants 2P03C00511p01
and 2P03C00511p04.
}

\bsk
\baselineskip = 12pt

% beginning of font "references"

{\references \ni REFERENCES
\ssk

\ref Bulik, T., Rudak, B., 1998, in preparation
\ref Camillo F., Thorsett S.E., Kulkarni S.R., 1994, ApJ, 421, L15
\ref Daugherty J.K., Harding A.K., 1989, ApJ, 336, 861
\ref Dyks J., Rudak B., 1998, these proceedings
\ref Edelstein J., Foster R.S., Bowyer S., 1995, ApJ, 454, 442
\ref Nel H.I. et al., 1996, ApJ, 465, 898 
\ref Rudak B., Dyks J., 1998a, MNRAS, 295, 337 
\ref Rudak B., Dyks J., 1998b, MNRAS, submitted
\ref Sturner S.J., Dermer C.D., 1994, A\&A, 281, L101 
}

\end{document}